\documentstyle[prl,aps,epsf,float,times]{revtex}
\begin{document}
\draft
\twocolumn[\hsize\textwidth\columnwidth\hsize\csname@twocolumnfalse\endcsname
\title{Theory of Scanning Tunneling Spectroscopy of
Magnetic-Field-Induced Discrete Nodal States in a D-Wave 
Superconductor}
\author{Boldizs\'ar Jank\'o}
\address{Materials Science Division, Argonne National Laboratory, 9700
  South Cass Avenue, Argonne, Illinois 60439} 
\date{December 23, 1998} \maketitle
\begin{abstract}
  In the presence of an external magnetic field, the low lying
  elementary excitations of a d-wave superconductor have quantized
  energy and their momenta are locked near the node direction.  It is
  argued that these discrete states can most likely be detected by a
  local probe, such as a scanning tunneling microscope. The low
  temperature {\em local} tunneling conductance on the Wigner-Seitz
  cell boundaries of the vortex lattice is predicted to show peaks
  spaced as $\pm \sqrt{n},\, n =\{ 0,1,2, ...\}$. The $n=0$ peak is
  anomalous, and it is present only if the superconducting order
  parameter changes sign at certain points on the Fermi surface. Away
  from the cell boundary, where the superfluid velocity is nonzero,
  each peak splits, in general, into four peaks, corresponding to the
  number of nodes in the order parameter.
\end{abstract}
]
\makeatletter
\global\@specialpagefalse
\def\@oddhead{REV\TeX{} 3.0\hfill ANL-MSD Theory Group Preprint, 1998}
\let\@evenhead\@oddhead
\makeatother

The nature of the quasiparticle spectrum in the Abrikosov vortex state
of the cuprate superconductors has attracted strong continuing
interest in recent years\cite{palee}. There are two main reasons why
this spectrum is expected to be unconventional. First, the pair size
$\xi_0$ seems to be comparable to the interparticle distance $1/k_F$.
Indeed, angle-resolved photoemission experiments \cite{jcc} on
$\rm{Bi}_2\rm{Sr}_2\rm{Ca}\rm{Cu}_2\rm{O}_{8+x}$ give $k_F \sim 0.7 \,
\AA^{-1}$, while magnetization studies \cite{pomar} indicate, that
$\xi_0 \sim 10-15\, \AA$. Second, there is by now substantial
experimental evidence \cite{nodes} for nodes in the cuprate
superconducting gap.  In contrast to conventional superconductors,
where the main source of low lying excitations in the mixed state are
the bound vortex core states \cite{caroli}, the above mentioned
reasons turn the bound core states into a delicate
feature\cite{maki,ichioka,franz} of cuprate superconductors.
Experiments seem to mirror this uncertainty: while bound core states
were observed in $\rm{Y}\rm{Ba}_2\rm{Cu}_3\rm{O}_{6-\delta}$ by
far-infrared spectroscopy\cite{drew} and scanning tunneling
spectroscopy (STS)\cite{maggio}, they seem to be absent in
$\rm{Bi}_2\rm{Sr}_2\rm{Ca}\rm{Cu}_2\rm{O}_{8+x}$ \cite{renner}.

The aim of the present paper is to point out, that a discrete spectrum
should nevertheless be observable in the {\em local} density of states
of the mixed state of all cuprates. It is somewhat unexpected, however,
that such a spectrum is more likely to be observable close to the
Wigner-Seitz cell boundaries of the Abrikosov lattice, where the
superfluid velocity is small. This is in contrast to conventional
Caroli--de Gennes--Matricon bound states\cite{caroli}, which appear in
the vortex core, in the region of singular superflow.  STS is an ideal
probe\cite{ali} to detect these low lying discrete states,
provided that the intervortex region is studied, and not the
core region, where the recent STS studies \cite{maggio,renner} were
focused so far.
The existence of field-induced discrete quasiparticle states in a
superfluid with gap nodes was suggested by Volovik \cite{volovik} for
the A phase of superfluid $^3{\rm He}$, where an effective magnetic
field is supplied by a static order parameter texture.  The analogue
of these nodal states in cuprate superconductors was recently proposed
by Gor'kov and Schrieffer\cite{gorkov}, as the appropriate basis
states for discussing the de Haas--van Alphen oscillations in the
mixed state of a d--wave superconductor. Quite recently, Anderson
\cite{anderson} attributed the anomalous magnetothermal conductivity
of $\rm{Bi}_2\rm{Sr}_2\rm{Ca}\rm{Cu}_2\rm{O}_{8+x}$ \cite{ong} to the
appearance of the magnetic field--induced discrete spectrum at the gap
nodes. These global probes, however, cannot provide direct
spectroscopic evidence for such a fine structure in the electron
spectrum: spatial averaging results in averaging over large Doppler
shifts \cite{cyrot} due to the supercurrents surrounding the vortices,
which in turn act to smear out the discrete level spectrum
\cite{average}.

\begin{figure}
\centerline{\epsfxsize=3.0in\epsffile{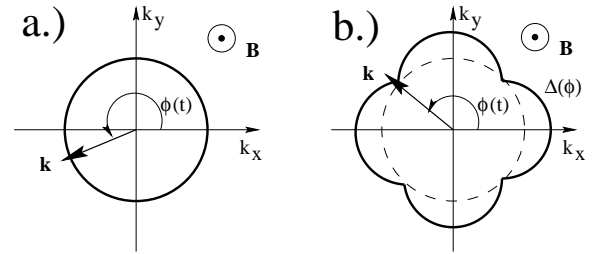}}
\vspace*{1ex}
\caption{(a) Schematic picture of a normal state quasiparticle 
momentum precessing according to $\phi (t) = \omega_c t$; (b) in the
presence of an anisotropic order parameter, the quasiparticle momenta 
of the low-lying excitations are bound to the node region $\phi_j =
(2j+1)\pi/4, \, j=\{0,1,2,3\}$.}
\label{fig:fs}
\end{figure}
One can anticipate the existence of such a discrete spectrum by the
following quasiclassical argument. Let us take a two dimensional
circular Fermi surface for the quasiparticles in the (a,b) plane and
and orient the magnetic field $\bf B$ along the c axis 
[see Fig.~\ref{fig:fs}(a)].  The quasiparticle equation of motion in 
the normal state is
\begin{equation}
\partial_t {\bf k } \; = \; \frac{e}{c\hbar} {\bf v_{\bf k}} \times {\bf B} -
\frac{\bf k}{\tau},
\end{equation}
where ${\bf v_{\bf k}} =\hbar^{-1} \partial_{\bf k} \epsilon (\bf k)$
is the quasiparticle velocity. While the normal state quasiparticles
are strongly damped, the quasiparticle lifetime $\tau$ increases
rapidly below the superconducting transition \cite{norman}. If that
would be the only change for temperatures below $T_c$, the solution to
the above equation could be given in terms of a phase variable
$ {\bf k} = k \, (\cos\phi(t), \sin \phi(t))$,
\begin{equation}
\phi (t) \; = \; \omega_c \frac{c\hbar}{eB}\int_{FS} 
\frac{dk}{v_k} = \omega_c t,
\end{equation}
where $\omega_c = (eB)/(m^*c)$.
More importantly, however, the onset of superconductivity generates
the appearance of a gap tied to the entire
Fermi circle [see Fig.~\ref{fig:fs}(b)]. If $\Delta (\phi) $ is
anisotropic, such as in the case of a d-wave superconductor with
$\Delta (\phi) = \Delta_0 \cos (2 \phi)$, the order parameter will 
provide an off-diagonal confining potential in the angular 
coordinate space, for the Bogoliubov quasiparticles precessing 
around the Fermi surface.

In order to provide a more detailed discussion of these phenomena, it
is necessary to obtain the quasiparticle amplitudes
$u_n(\phi),v_n(\phi)$. Let us focus on the intermediate field regime
$B_{\rm c1} \ll B \ll B_{\rm c2}$ when the field in the sample is
fairly uniform, and address the effect of superflow later.
It is most convenient to use the Bogoliubov--de Gennes equations for
the angular amplitudes of quasiparticles near the Fermi surface $k \sim
k_F$, in the form given by Gor'kov and Schrieffer
\cite{gorkov,others}:
\begin{eqnarray}
(E_n - i\hbar\omega_c \partial_\phi + \bar\mu)u_n(\phi) + \Delta(\phi)
v_n(\phi) &\; = \;& 0, \nonumber\\
(E_n + i\hbar\omega_c \partial_\phi - \bar\mu)v_n(\phi) + \Delta(\phi)
u_n(\phi) &\; = \;& 0.
\label{eq:andreev}
\end{eqnarray}
where $\bar\mu$ is defined by $ \mu - \bar\mu = \omega_c N_0$, ($N_0$
is a large integer). In this approximation, the dependence of the gap
and the amplitudes on the radial component of the momentum is
neglected, and the angle $\phi $ remains the only dynamic variable.
The amplitudes obey periodic boundary conditions $
(u_n(\phi),v_n(\phi)) = (u_n(\phi + 2\pi ),v_n(\phi + 2\pi))$.  The
normalization conditions are
\begin{equation}
\label{eq:norm}
\int_{-\pi}^{\pi} \frac{d\phi}{2\pi}[ |u_n(\phi)|^2 + |v_n(\phi)|^2 ]
\; = \; 1.
\end{equation}
As will soon become evident [cf. Eq.(\ref{eq:spectrum})], 
$E_n \gg \bar\mu$, and therefore
$\bar\mu $ will be neglected in the following calculations. 
Using the observation of Bar-Sagi and Kuper\cite{barsagi}, and Kosztin
and coworkers\cite{kosztin} for the Andreev Hamiltonian, 
the {\em square} of the Hamilton operator acting on $
(u_n(\phi),v_n(\phi))$ in (\ref{eq:andreev}),
can be diagonalized with the help of a unitary transformation. 
The eigenvalues and the eigenfunctions of eigenvalue problems obtained
in this way
  \begin{eqnarray}
[-\hbar^2\,\omega_c^2\,\partial^2_\phi  + \Delta^2(\phi) \pm
  \hbar\,\omega_c\, \left(\partial_\phi \Delta (\phi)\right)]
\,\Phi_{\pm,n} (\phi) \nonumber\\
 \;=\;  \lambda_n \,\Phi_{\pm,n} (\phi)\;.
    \label{eq:eigen}
  \end{eqnarray}
are simply related to the eigenvalues and  eigenfunctions, 
$E_n, (u_n(\phi),v_n(\phi))$, of the original problem
(\ref{eq:andreev}):
  \begin{eqnarray}
   \label{eq:uv}
   E_n &\;=\;& \pm \sqrt{\lambda_n}\nonumber\\
    u_n(\phi) &\;=\;&
\frac{1}{2}[\Phi_{-,n}(\phi) + i \Phi_{+,n}(\phi)] \;, \nonumber\\
    v_n(\phi) &\;=\;&
\frac{1}{2i}[\Phi_{-,n}(\phi) - i \Phi_{+,n}(\phi)] \;.
  \end{eqnarray}
Furthermore, the eigenfunctions of the two branches are interrelated
for $\left|E_n\right| > 0$ :
  \begin{equation}
    \label{eq:ladder}
    \Phi_{-,n} = \frac{1}{\left|E_n\right|}\,Q\,\Phi_{+,n}\;, \quad
    \Phi_{+,n} = \frac{1}{\left|E_n\right|}\,Q^{\dagger}\,\Phi_{-,n}\;.
  \end{equation}
Here the following notation was used 
$Q \;\equiv\; -\hbar\,\omega_c\,\partial_{\phi } + \Delta\;, 
    Q^{\dagger} \;\equiv\; \hbar\,\omega_c\,\partial_{\phi } + \Delta\;$.
In the intermediate field regime the excitations of interest lie deep
in the node region $\theta = \phi - \phi_j \ll \phi_j$, for which the
d-wave potential is linear $ \Delta (\phi) \simeq 2 \Delta_0
\theta$. Thus, near the nodes, Eq.(\ref{eq:eigen}) takes a
particularly simple form 
\begin{equation}
\Psi''_{-,n} (\theta) + \Bigg[\frac{E_n^2 + 2\hbar\omega_c
  \Delta_0}{\hbar\omega_c^2} 
- 2\frac{\Delta_0}{\hbar\omega_c}\theta^2\Bigg]\Psi_{-,n} (\theta) \; = \; 0,
\end{equation}
which can be recognized as the Schr\"odinger equation of a simple
harmonic oscillator, with solutions proportional to $\exp[-
(\Delta_0/\hbar\omega_c) \theta^2]$. For typical fields $B = 9 \, T$, 
and gap values \cite{jcc} $ \Delta_0 \simeq 40\,{\rm meV}$, the
ratio $ \gamma \equiv 2\Delta_0/(\hbar\omega_c) \sim 80 $. This means that the 
significant weight of these states is strongly localized around the
node regions. Extending the limits of integration to infinity in the
normalization condition (\ref{eq:norm}), we obtain for the eigenfunctions
\begin{equation}
\label{eq:hermite}
\Psi _{-,n} (\theta ) \; = \; c_n H_n [\gamma^{1/2} \theta ] 
e^{-\frac{\gamma}{2} \theta^2}
\end{equation}
where $H_n(x)$ are the Hermite polynomials, and the coefficient $c_n$ is 
\begin{equation}
c_n = \sqrt{\frac{(\gamma/\pi)^{1/2}}{2^n\,n!\,[(n+2) \, + \, (5\gamma/4n) \, +
\, (2\gamma)^{1/2}]}}, \quad n > 0.
\end{equation}
The angular amplitudes obeying the periodic boundary conditions can be
constructed from the above solution, Eq.(\ref{eq:hermite}), by
inserting them into eqs.(\ref{eq:uv}) and (\ref{eq:ladder}) with the
substitution $\Phi_{-,n} (\phi) = \Psi [min_j\{\phi -(2j+1)\pi/4\}]$.
The overlap of the nodal states residing in two adjacent nodes is
$\exp[ - (\pi^2 \Delta_0)/(8\hbar\omega_c)] \sim 3.7 \, \times \,
10^{-22}$. Clearly, quantum oscillations arising from
interference between low lying nodal states located at different nodes
will be significant only near $H_{c2}$, where $ \hbar \omega_c \sim
\Delta_0$. The eigenvalues are $E_n^2 = 4 n \Delta_0 \hbar\omega_c$,
which gives two branches of eigenvalues for the original equations
(\ref{eq:andreev})
\begin{equation}
\label{eq:spectrum}
E_n = \pm 2 \sqrt{n \Delta_0\hbar\omega_c}
\end{equation}
The spacing between these states is considerably larger than the
magnetic energy $\hbar \omega_c \ll 2 \sqrt{ \Delta_0\hbar\omega_c}
\ll \Delta_0$. For $B = 9 \, {\rm T}$,  
$\delta E_n = (2\gamma )^{1/2} \hbar\omega_c \sim 13 \, {\rm meV}$, while
$\hbar\omega_c \simeq 1\, {\rm meV}$. 

The $n=0$ state is anomalous, as it does not belong to either of the
sets in (\ref{eq:ladder}). It is a so-called ``zero-mode'' a feature
that is expected for the supersymmetric {\em squared} Hamiltonian 
\cite{witten,kosztin}.  It is worth noting that even a strongly
anisotropic s-wave order parameter could, in principle, give a
discrete quasiparticle spectrum in a magnetic field. However, the
conditions of normalizability required for the existence of such a 
state are naturally satisfied {\em only} for 
an order parameter with true gap nodes [ such that $\Delta (\phi)
\sim [\partial_{\phi} \Delta (\phi)]_{\phi = \phi_j} (\phi - \phi_j)$ 
near $\phi \sim \phi_j$ ]. 

Clearly, the experimental detection of the discrete nodal spectrum and
especially, the observation of the anomalous zero mode in materials
suspected of having gap nodes, would be quite important.  Let us now
find the most favorable experimental circumstances for observing these
states. As in the case of normal state Landau levels, the
semiclassical trajectories associated with the discrete nodal states
are orbits of radii of the order of the magnetic length $d =
\sqrt{2eB/(hc)}$ \cite{anderson} where $B$ is the applied magnetic field, and $
hc/(2e)$ is the superconducting flux quantum.  An external field of $B
= 9 \,{\text T}$ gives $d \simeq 150 \, \AA$. This length also gives
the lattice constant of the Abrikosov vortex lattice: $a_{\Box} = d;
\, a_{\triangle} = (4/3)^{1/4} \, d$ \cite{abrikosov}. Since the
motion occurs on the background of a strongly varying superflow the
spectrum is shifted \cite{cyrot} by $ \delta E_S (r) = m\,{\bf v}_F \,
\cdot \, {\bf v}_s(r)$ typically of order $ \delta E_S \sim (\Delta_0
\xi_0)/(2d)$. This is {\em not} negligible when compared to the level
spacing $\delta E_n \sim 2 \sqrt{\hbar\omega_c \Delta_0} \sim 2
\Delta_0 \sqrt{\ell_F\xi_0}/d$ [here $\ell_F = 2\pi/k_F \sim 0.23 \,
\AA$ is the Fermi wavelength]. Indeed,
\begin{equation}
\frac{\delta E_n}{\delta E_S} \sim 4 \,
\Bigg(\frac{\ell_F}{\xi_0}\Bigg)^{1/2} \sim 0.5
\end{equation}
If the trajectory includes segments close to the core $r \sim \xi_0$, in
those regions the ratio drops to $\delta E_n/ \delta E_S \sim 0.05$. A
global spectroscopic probe, such as electromagnetic absorption
\cite{drew}, will detect a spatially averaged response. Clearly, 
this also implies an average over the shift $\delta E_S$,  which
would make the resonant transitions between the discrete quasiparticle
levels difficult to observe.

The situation is quite different for a scanning tunneling microscope,
with which one could select with at least $\delta r \sim 1\, \AA$
resolution the place where electrons are injected or removed from
a superconductor. The local tunneling conductance can be given as
\begin{eqnarray}
\label{eq:cond}
\Big(\frac{d I(r)}{dV} \Big) \propto - N_B \sum_{n \ge \, 0} \int^{\pi}_{-\pi}
\frac{d\phi}{2\pi} |T(\phi)|^2 \times \nonumber \\
\{|u_n(\phi)|^2 \frac{\partial f[E^{-}_n(r,\phi,V)]}{\partial E_n}
+ |v_n(\phi)|^2 \frac{\partial f[E^{+}_n(r,\phi,V)]}{\partial E_n}
\}. 
\end{eqnarray}
Here $N_B = ({\rm sample \, area})/(2\pi d^2)$ is the degeneracy of each
level, $|T(\phi)|^2$ is the square of the tunneling matrix element which,
in principle, can be angle dependent \cite{directional}, 
$E^{\pm}_n(r,\phi,V) = E_n(r,\phi) \pm eV$ and
$f(\epsilon) = [\exp(\beta\epsilon) + 1]^{-1}$ is the Fermi
distribution function. 
The local spectrum also depends on the angle $\phi $ through the Doppler shift
\begin{equation}
  \label{eq:localspectr}
E_n(r,\phi) =  E_n - E_S(r) \cos(\alpha + \phi),
\end{equation}
where $\alpha $ is the angle between the local superfluid velocity and
the $\hat a$ crystal axis. For a rigid and stationary Abrikosov
lattice $\alpha $ is fixed. $E_S(r) = m v_F v_s(r)$ is the maximum
Doppler shift at particular place ${\bf r}$.  In Eq.(\ref{eq:cond})
there is no summation with respect to $k_z$, the momentum in the $\hat
c$ direction. This two dimensional approximation is especially
adequate for superconducting
$\rm{Bi}_2\rm{Sr}_2\rm{Ca}\rm{Cu}_2\rm{O}_{8+x}$ samples
\cite{kleiner}.

\begin{figure}
\vspace*{-0.5cm}
\leftline{\hspace*{0.1in}\epsfxsize=3.5in\epsffile{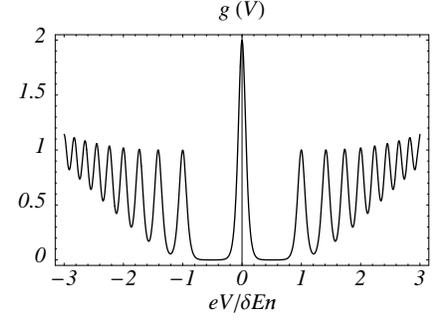}}
\vspace*{-1.5cm}
\caption{The dimensionless STS tunneling spectrum $g(V)$ on the
Wigner-Seitz cell boundary, according to Eq.(\ref{eq:cond2}),
for $k_BT/\delta E_n  = 1/24$.}
\label{fig:spectrum}
\end{figure}
\begin{figure}
\vspace*{0.4in}
\leftline{\epsfxsize=3in\epsffile{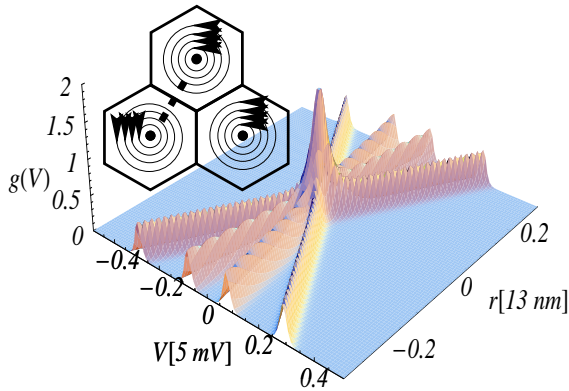}}
\vspace*{-2truecm}
\caption{
  {\bf Main Figure:} Plot of the dimensionless conductance $g(V)$ [cf.
  Eq.(\ref{eq:cond2})] of the {\em zero bias peak only}, as the
  position is varied along the dotted line of the inset. The
  parameters are $\delta E_n = 5 \, {\rm meV}, d = 400 \, \AA$, $
  \alpha = \pi/6$, $T = 6 \, {\rm K}$.  {\bf Inset}: Schematic figure
  of a vortex lattice. The Wigner-Seitz cells, where the superfluid
  velocity is zero, are drawn with heavy lines; the dots correspond to
  the vortex cores (not on scale with the lattice constant). The
  suggested path for the tip of the scanning tunneling microscope is
  shown with a heavy dotted line. }
\label{fig:wigner}
\end{figure}

The important observation here is that according to
Eq.(\ref{eq:hermite}) the Bogoliubov amplitudes $u_n(\phi), v_n(\phi)$
are highly peaked functions around the nodal angles.  Compared to 
$|u_n(\phi)|^2$ and $|v_n(\phi)|^2$, all other factors in
(\ref{eq:cond}) are weak functions of the angle $\phi$ , and can be taken
outside of the integral.  As a consequence the integration over the
angle reduces to summation over the nodal directions of the order
parameter: For a d-wave superconductor, $\phi_j = (2j+1)\pi/4, \,
j=\{0,1,2,3\}$.  While the tunneling matrix element along the node
directions $|T(\phi_j)|^2$ is small \cite{directional}, it is
nevertheless finite, as indicated by the finite zero bias conductance
in tunneling experiments \cite{renner}. In the following, it is
assumed that $|T(\phi_j)|^2$ has the same value for all four nodes.
The dimensionless conductance $g(V)$ can be defined as
\begin{eqnarray}
\label{eq:cond2}
g(V) \equiv - \sum_{j,n \ge \, 0} \Bigg\lbrace
\frac{\partial f [E^{-}_n(r,\phi_j,V)]}{\partial (\beta E_n) }
+ \frac{\partial f[E^{+}_n(r,\phi_j,V)]}{\partial (\beta E_n)} 
\Bigg\rbrace. 
\nonumber\\
\end{eqnarray}
In order to reveal the essential features of the above result, let us
first assume that the tip of the scanning tunneling microscope is
positioned on the Wigner-Seitz cell of the vortex lattice (cf. the
inset of Fig.\ref{fig:wigner}). For all point on the cell boundary
$E_S(r) =0$. At low enough temperature the quasiparticle scattering
rate \cite{bonn}, as well as the uncertainty in $\delta E_S(r)$ due
to vortex lattice fluctuations is smaller than the level spacing $
\delta E_n > \hbar/\tau_{\rm qp}, \delta E_S^{\rm lattice}$.  Under
these conditions the tunneling conductance reveals (see
Fig.\ref{fig:spectrum}) the discrete spectrum obtained in
Eq.(\ref{eq:spectrum}).
Note that the amplitude of the zero mode is twice the amplitude of the
finite voltage peak, since the particle and hole tunneling rates are
the same near $V=0$.  This is also evident from Eq.(\ref{eq:cond}).

As the tip is moved away from the cell boundary, 
towards one of the cores, (for example, along the dashed line of inset
of Fig.\ref{fig:wigner}), each peak splits up into four
separate peaks, corresponding to the four nodes of the order
parameter. This is illustrated in the main part of
Fig.\ref{fig:wigner}, where, for clarity, only the splitting of
the zero bias peak is shown.  The amplitudes of the split peaks are 
four times smaller than the amplitude of the single peak measured at the
Wigner-Seitz cell boundary. 

In conclusion, this paper addressed the problem of the quasiparticle
spectrum in a d-wave superconductor in an external magnetic field. The
spectrum is discrete, and the quasiparticle amplitudes are strongly
peaked for momenta pointing along the node directions. An anomalous
zero mode is present only if the order parameter has true nodes.  As
it turns out, the level spacing is smaller than the average Doppler
shift due to the interaction of quasiparticles with the supercurrents
circling around the vortices. Thus, in order to detect this spectrum,
one needs to perform scanning tunneling spectroscopy on the
Wigner-Seitz cell boundary of the Abrikosov vortex lattice, where the
superfluid velocity is zero.  The calculated local tunneling spectrum
also indicates that slightly away from the Wigner-Seitz boundary, each
peak splits into four subpeaks, as a consequence of the four different
Doppler shifts arising for quasiparticles tunneling into the four
nodes. The results are clearly more general than the specific case of
a quasi-two dimensional superconductor with a d-wave order parameter,
and can be extended in a straightforward manner to {\em any layered
  superconductor that has gap nodes}. In fact, the experiment proposed
here would probe intimate details of the superconducting state, such
as the {\em existence}, the {\em number} and the {\em position} of gap
nodes on the Fermi surface.

I would like to thank Prof. Alexei A. Abrikosov and 
Dr. Ioan Kosztin for helpful discussions.
This research was supported in part by the NSF under awards
DMR91-20000 (administrated through the Science and Technology Center
for Superconductivity), and the U.S.  DOE, BES, under Contract
No.~W-31-109-ENG-38.

\end{document}